# PLASMONIC NANOPARTICLE-BASED PROTEIN DETECTION BY OPTICAL SHIFT OF A RESONANT MICROCAVITY

Miguel A. Santiago-Cordoba[1], Svetlana V. Boriskina[2], Frank Vollmer[*,3,4], Melik C. Demirel[*,1,3]

1. Department of Engineering Science, and Materials Research Institute, Pennsylvania State University, University Park, Pennsylvania, 16802, USA.
2. Department of Chemistry and Electrical and Computer Engineering, Boston University, Boston, Massachusetts, 02215, USA
3. Wyss Institute for Bioinspired Engineering, Harvard, Cambridge, Massachusetts 02138, USA.
4. Current address: Max Planck Institute for the Science of Light, Laboratory of Biophotonics & Biosensing, 91058 Erlangen, Germany.

* Correspondence should be addressed to MCD and FV. Email: mdemirel@engr.psu.edu, frank.vollmer@mpl.mpg.de

**ABSTRACT**

We demonstrated a biosensing approach which, for the first time, combines the high-sensitivity of whispering gallery modes (WGM) with a metallic nanoparticle based assay. We provided a computational model based on generalized Mie theory to explain the higher sensitivity of protein detection through Plasmon enhancement. We quantitatively analyzed the binding of a model protein (i.e., BSA) to gold nanoparticles from high-Q WGM resonance frequency shifts, and fit the results to an adsorption isotherm, which agrees with the theoretical predictions of a two-component adsorption model.

A whispering-gallery-mode (WGM) biosensor uses high-Q optical resonances to directly detect the binding of molecules and nanoparticles from a frequency shift signal[1-4]. The WGM biosensing approach is highly sensitive, down to the single virus and nanoparticle level[5, 6]. However, rapid sample analysis (given the small sensing area), varied surface functionalization of monolithic sensor arrays, as well as specific detection against a complex backround are some of the challenges in this field[7, 8]. Nanoparticle (NP) biosensing platforms such as those based on gold (Au) NPs, on the other hand, may overcome some of these challenges by providing a large sensing area, rapid mixing of the analyte, and the possibility for extracting NPs from a complex sample for subsequent optical analysis[9-11].

In this letter, we combine the advantages of a NP-based assay with the high sensitivity of a WGM biosensor to determine the amount of protein bound to NPs by analyzing frequency/wavelength shifts of hybrid photonic-plasmonic modes. We use Au NPs to nonspecifically adsorb Bovine Serum Albumin (BSA) protein from a phosphate buffered saline (PBS) solution. After incubation in the BSA solution (SeraCare Life Sciences, Inc), the NPs are filtered, and immobilized on an anodic aluminum oxide (AAO) capture-membrane. Subsequently, the dry NP layer is interrogated by observing high-Q WGM resonance frequency shifts that occur after evanescent coupling of a microsphere cavity to NPs in the layer. By analyzing WGM resonance frequency shifts for NPs exposed to different BSA concentration levels, we find that the adsorption isotherm agrees with a two-component adsorption model. We show picomolar sensitivity levels for this approach which decouples WGM transducer from (NP) recognition element.

The method for preparation of the Au-NP solution with BSA, and capture of the NP layer on the AAO membrane (AAO, Whatman Inc.) by vacuum suction are shown schematically in Figure 1a. The preparation of the Au NPs was done following a modification of the Lee and Meisel's method as described by Kruszewski *et al.*[12] Figure 1b shows the top surface electron micrograph of bare AAO substrate and the Au NPs captured on the AAO membrane.

The resonant microsphere cavity structure and its orientation with respect to the Au NPs captured on the AAO membrane are illustrated in Figure 1c. We used a well established WGM setup that monitors WGM resonances of a tapered-fiber-coupled microsphere in real-time[13]. We determined wavelengths shifts associated with coupling of WGM to the NPs by recording and subtracting WGM resonance wavelengths measured for the cavity held in air, and then in contact with the NP layer. Typical WGM spectra before and after coupling to Au NP layer is shown in Figure 1d. We record the shift for two WGMs centered around ~632.014 nm (Peak 1) and ~632.016 nm (Peak 2) wavelengths.

The silica microsphere (diameter of 450±10 μm) probes the NP layer via its evanescent field which extends about $L$~50 nm at 633 nm wavelength (L~80 nm at 1060 nm wavelength) outward from the microsphere surface, where $L = \lambda[4\pi(n^2_{eff,sphere} - n^2_{eff,medium})^{1/2}]^{-1}$, $n_{eff,sphere}$ ~1.45 and $n_{eff,medium}$ ~1 are the effective refractive indices of the glass microsphere and that of the medium surrounding the NPs (air), and $\lambda$ is the nominal wavelength of the laser[14]. The limited extend of L ensures that we are probing only the first layer of Au-NPs. Typical Q factors before coupling the microsphere to Au-NP are ~ $3\times10^6$ at 633nm and ~$5\times10^6$ at 1060 nm.

Next we compare the Au NP WGM shift at ~ 633 nm to that at ~ 1060 nm wavelength to test for signal enhancement due to plasmon coupling. For the same Au NP layer the wavelength shift averaged over different probing locations is larger at ~ 633 nm wavelength ($\Delta\lambda = 1.2 \pm 0.1 \times 10^{-4}$ nm) as compared to ~1060 nm wavelength ($\Delta\lambda = 0.8 \pm 0.1 \times 10^{-4}$ nm). Coupling WGM to NP plasmon resonance (~525 nm) can thus enhance signal to noise ratio (SNR).

We performed exact numerical calculations based on the generalized multi-particle Mie theory to reveal the detailed picture of the WGM interaction with the localized surface plasmon (LSP) resonance of the Au NP[15, 16]. Experimentally obtained Au refractive index values were used in the simulations[17]. We selected a smaller sphere (5μm) for the simulation to limit the computation time. Figure 2a shows the simulated values of WGM shifts, $\Delta\lambda/\lambda$, for a single NP bound at the center of the evanescent tail. The hybrid photonic-plasmonic modes due to the coupling of the narrowband WGM with the plasmon resonance feature significant electric field enhancement on the NP surface (Figure 2b) and can be efficiently excited in a wide frequency range covering the longer-wavelength slope of the NP LSP resonance peak[15, 16]. Indeed, our calculations in Figure 2b show up to 3 orders-of-magnitude increase of the electric field intensity on the NP surface. Strong WGM coupling to the NP collective electron oscillations at the frequencies close to the NP dipole LSP resonance increases both radiative and dissipative losses, resulting in the decrease of the Q-factors of hybrid modes in the 520-580 nm frequency range (Figure 2c). This decrease of the structure's ability to accumulate energy from the external excitation field is reflected in the drop of the field intensity close to the NP LSP resonance observed in Figure 2b.

The spatial distributions of the local field intensity near the surface of the microsphere at the resonance wavelength λ~632 nm in the presence and absence of the Au NP are shown in Figure 2d and 2e respectively. In the frame of the first-order perturbation approximation[5], fractional wavelength shift of the mode caused by a small protein molecule with a real excess polarizability $\alpha$ at position $\mathbf{r}_v$ is directly proportional to the field intensity value at the molecule position $|\mathbf{E}(\mathbf{r}_v)|^2$ and inversely proportional to the energy density integrated over the whole mode volume.

$$\left(\frac{\Delta\lambda_r}{\lambda_r}\right) \cong \frac{\alpha/\varepsilon_0 |\mathbf{E}(\mathbf{r}_v)|^2}{2\int_V \varepsilon_r(\mathbf{r})|\mathbf{E}(\mathbf{r})|^2 dV} \qquad (1)$$

Significant increase of the field in the area accessible by the protein molecules in the microsphere-NP structure (compare Figure 2d and 2e) is expected to translate into the larger sensitivity of the hybrid WGM-NP sensor. Indeed, our calculations show that adsorption of a single BSA molecule (modeled as a 3.4 nm-radius nanosphere with *n* = 1.45) at the center of the evanescent field does not produce a detectable shift of the WGM resonance at λ ~ 580nm. However, the BSA placed in the hot spot created on the WGM-coupled NP causes Δλ = 1.06·10$^{-4}$ nm shift of the corresponding hybrid resonance. If the BSA molecule were instead placed in the hot spot of the isolated NP alone, it causes Δλ = 1.497·10$^{-2}$ nm shift of the dipole LSP resonance, due to the small NP mode volume which decreases the value of the denominator in equation (1). However, orders-of-magnitude larger Q-factors of hybrid photonic-plasmonic modes (Q ~ 10$^{6-7}$) in the WGM-NP structure over that of the single NP plasmon resonance increase the hybrid's sensor spectral resolution and greatly improves the BSA detection limit as compared to individual, WGM-based or NP-based sensors.

We tested our method with the detection of BSA protein from solution. We incubated NPs with different concentrations of BSA solutions ranging from pM to μM. The Au NPs adsorb BSA protein and after incubation for about three hours, the Au NP-BSA layer is formed on the AAO membrane for subsequent analysis with WGMs. In Figure 3, the WGM wavelength shifts for the two modes (peak 1 and 2, Figure 1d) averaged for different probing locations are plotted. From each measurement we subtracted the wavelength shift for a NP layer exposed to control solution which contains no BSA protein. For BSA concentration higher than 1 μM, we observed additional increase of the shift signal possibly due to BSA multilayer formation and solution aggregation (Figure 3 inset). We observed slight variation of the shift signal for peak 1 and peak 2 indicating possible influence of the WGM mode number and the polarization on the shift signal.

The isotherm in Figure 3 can be explained by a two-component adsorption model [18]. For each NP subsystem which has $m$ equivalent and independent sites for adsorption, the partition function pertaining to the single NP is $\xi = (1+q\phi)^m$, where $q$ is the NP concentration, $\phi = e^{\mu/kT}$, k is the Boltzmann constant, T is the Temperature, μ is the binding energy. The average number $\overline{N}$ of bound molecules in the macroscopic two-component system can be estimated from the ensemble of NP subsystems as,

$$\overline{N} = M\phi\left(\frac{\partial \ln \xi}{\partial \phi}\right)_T \text{ or } \overline{s} = \frac{\overline{N}}{M} = \phi\left(\frac{\partial \ln \xi}{\partial \phi}\right)_T \quad (2)$$

where M is the total number of available sites, s is the average number of adsorbed molecules per NP subsystem with $0 \leq s \leq m$. Inserting the partition function for the NP subsystem in equation 2 yields,

$$\overline{s} = \phi\left(\frac{\partial}{\partial \phi}\right)(m \ln(1+q\phi)) = m\phi\frac{q}{1+q\phi} = m\frac{q\phi}{1+q\phi} \quad (3)$$

For a single NP or interface (m=1), the equation 3 reduces to the classic Langmuir adsorption equation.

In conclusion, we demonstrated that WGM modes can be used to quantitatively probe the amount of BSA protein adsorbed to plasmonic Au NPs. To the best of our knowledge this is the first time that the high-sensitivity of a WGM biosensor is combined with a NP-based assay for optical detection. Future research will focus on increasing SNR by optimizing WGM and NP plasmon resonance, and studying spectroscopically[19-21] the binding curve between molecules and their NP-binding sites.

## ACKNOWLEDGMENT

We gratefully acknowledge financial support for this work from the Wyss Institute at Harvard, as well as the Rowland institute at Harvard for the cooperation in the use of their facilities.

## AUTHOR CONTRIBUTIONS

MCD and FV planned and supervised the research. MSC, FV and MCD performed the experiments. SVB performed the numerical simulations. All authors contributed to writing and revising the manuscript, and agreed on its final contents.

**FIGURE CAPTIONS**

**Figure 1.** a) Schematic of gold nanoparticle (Au NP) preparation. Au NPs were mixed with BSA and incubated at 4 °C for 3h to form BSA-Au NPs. AAO membranes (100 ± 10 nm hole diameter) were soaked for 5h in a ~5% solution of polyethyleneinimide (PEI) to enhance adhesion of Au NPs to AAO membrane. Subsequently, AAO membranes were washed with pure water, and dried in desiccators under vacuum. b) Electron microscopy images of bare AAO and AAO with Au NPs using FEI Philips XL-20 microscope. c) Schematic diagram of the set-up employed to evanescently couple WGMs to the Au nanoparticle layer. Top: WGM transmission spectrum for the microsphere in air. Bottom: WGM spectrum for microsphere in contact with nanoparticles which induce WGM wavelength shift. The microsphere cavity was fabricated by melting the tip of a single mode optical fiber. d) Example of a WGM spectrum in air (dotted line) and after evanescent coupling to a BSA-nanoparticle layer (continuous line).

**Figure 2.** (a) Spectral shifts of the first-radial-order WGM of a 5μm-diameter $SiO_2$ microsphere in air caused by the attachment of 55 nm-diameter Au NP (shown in the inset). (b) Intensity enhancement on the Au NP surface in the presence (red dots) and absence (blue line) of the sphere, in this case under the illumination by a linearly-polarized plane wave. (c) Q-factors of the resonances corresponding to the WGM in the microsphere and hybrid photonic-plasmonic modes in the coupled sphere-NP structure. Spatial electric field intensity distributions at the wavelengths of (d) a hybrid resonance (λ=630.84nm) and (e) the corresponding WG-mode resonance (λ=630.819nm) (sphere surface is shown as a dashed line).

**Figure 3.** WGM measurements at ~633 nm probing wavelength. Au NPs where incubated with different concentrations of a BSA solution and then immobilized onto the AAO membrane for WGM shift analysis. The inset shows the BSA adsorption at high solution concentrations compared to dilute concentration in a log-linear graph.

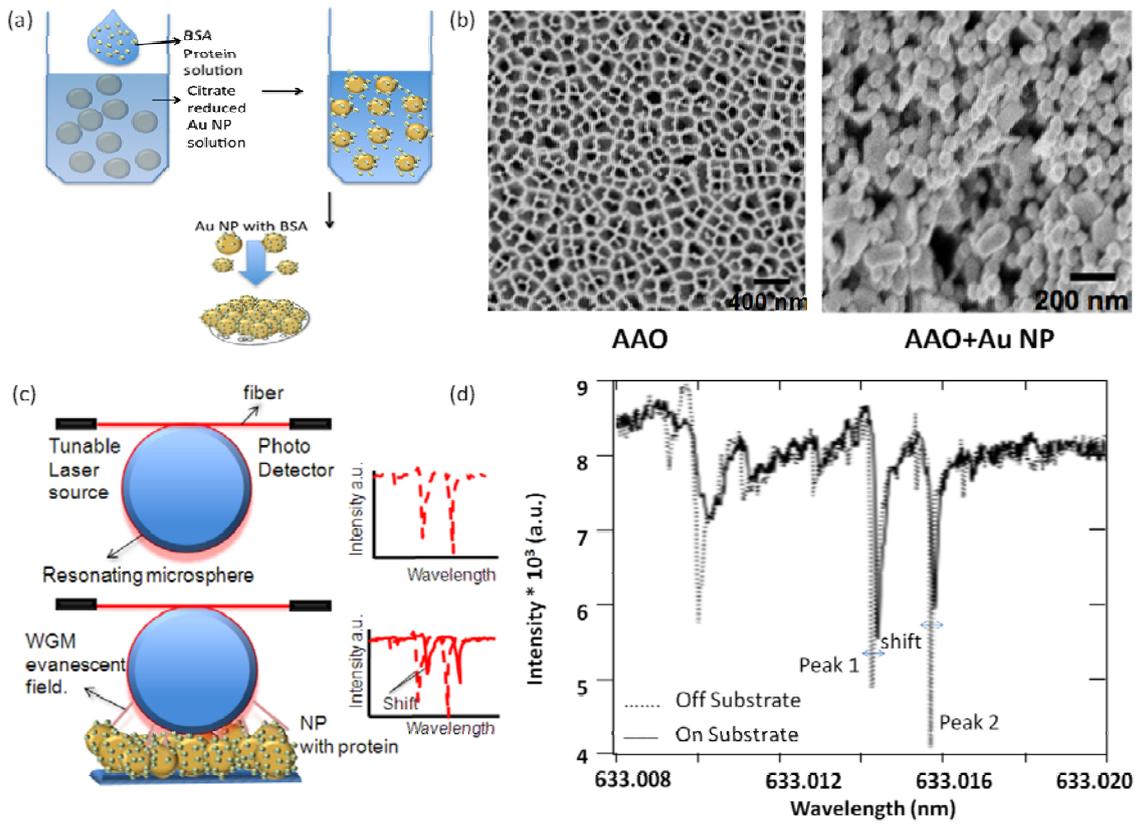

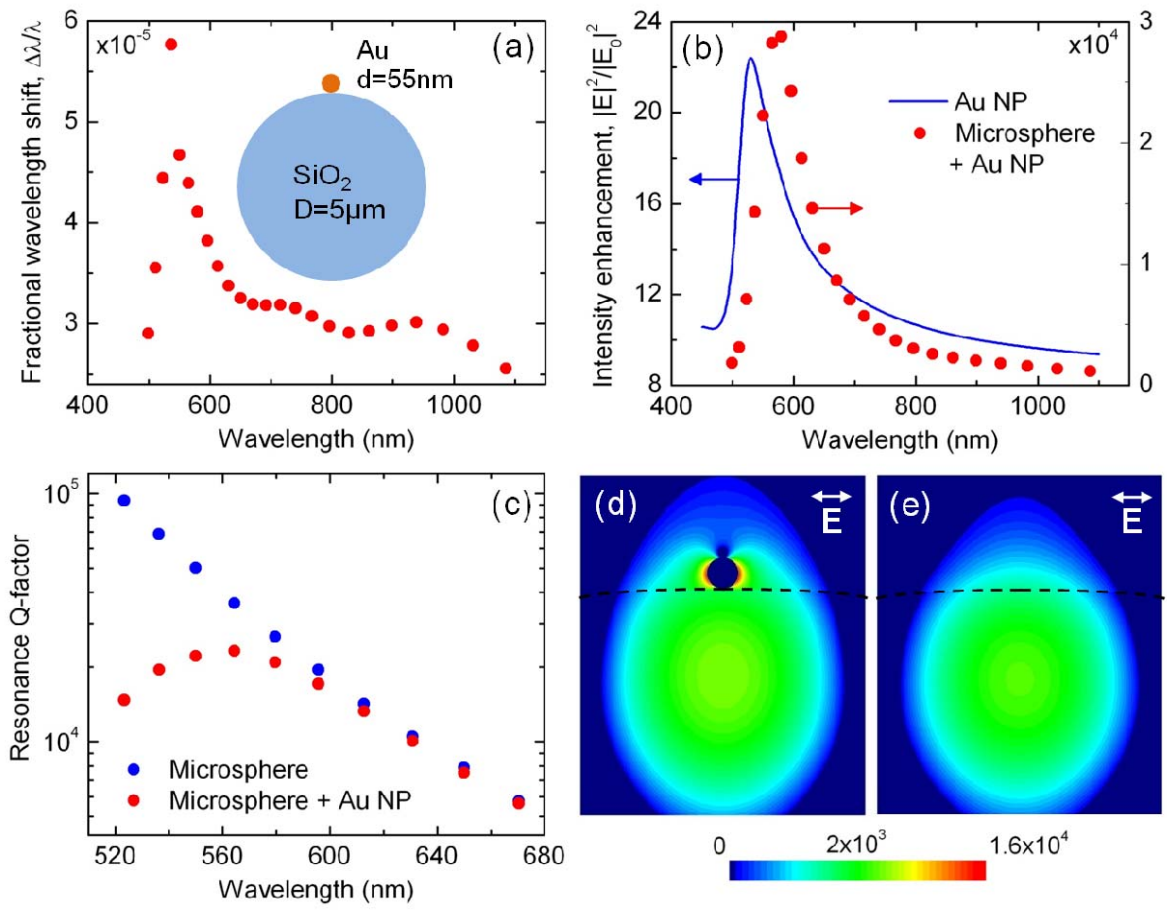

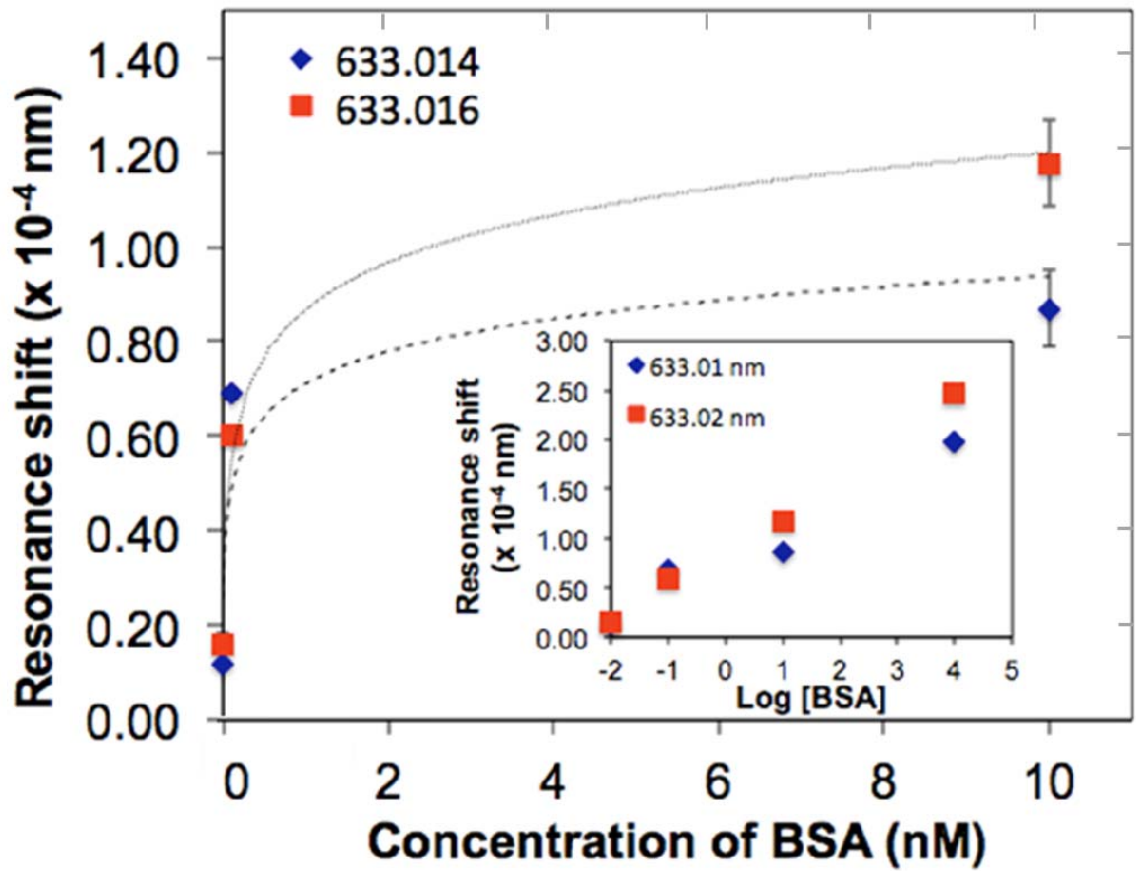